\documentclass[onecolumn,showpacs,preprintnumbers,amsmath,amssymb,showkeys,12ft]{revtex4}

\usepackage{graphicx}
\usepackage{dcolumn}
\usepackage{bm}

\begin{document}
\draft
\date{\today}
\title{Klimontovich's S theorem in nonextensive formalism and the problem of constraints}
\author{G. B. Ba\u{g}c\i}
\email{gbb0002@unt.edu@unt.edu}
\address {Department of Physics, University of North Texas, P.O. Box 311427, Denton, TX 76203-1427,
USA}

\pagenumbering{arabic}

\begin{abstract}
Ordinary Boltzmann-Gibbs entropy is inadequate to be used in systems
depending on a control parameter that yield different mean energy
values. Such systems fail to give the correct comparison between the
off-equilibrium and equilibrium entropy values. Klimontovich's S
theorem solves this problem by renormalizing energy and making use
of escort distributions. Since nonextensive thermostatistics is a
generalization of Boltzmann-Gibbs entropy, it too exhibits this same
deficiency. In order to remedy this, we present the nonextensive
generalization of Klimontovich's S theorem. We show that this
generalization requires the use of ordinary probability and the
associated relative entropy in addition to the renormalization of
energy. Lastly, we illustrate  the generalized S theorem for the Van
der Pol oscillator.
\end{abstract}

  \narrowtext
\newpage \setcounter{page}{1}

\maketitle

\section{\protect\bigskip Introduction}

\noindent \qquad

Recently, a new measure of complexity called renormalized entropy
(RE) has been proposed [1]. This measure is based on Klimontovich's
S theorem which states that the renormalized entropy decreases in
the process of self-organization [2-5]. Originally, the S theorem
was used by Klimontovich to remedy the failure of Boltzmann-Gibbs
(BG) entropy when it is used in open systems. In these cases, BG
entropy resulted out of equilibrium entropy values greater than the
corresponding equilibrium case. In order to solve this problem, he
made use of escort distributions and energy renormalization. By
equating the mean energy values, he was able to redefine the
intensity of random source in such a way that the equilibrium
entropy value was maximum once again. Klimontovich's S theorem has
been used in many fields ranging from logistic map [1], heart rate
variability [6, 7] to the analysis of electroencephalograms of
epilepsy patients [8]. Later, Quiroga et al. [9] have shown that RE
is negative of the Kullback-Leibler (KL) entropy [10] i.e., the
relative entropy associated with the ordinary Boltzmann-Gibbs (BG)
entropy once the reference distribution is taken to be the
renormalized escort distribution [11, 12, 13].

\noindent \qquad In this paper, we generalize Klimontovich's S
theorem and construct corresponding renormalized entropy measure in
nonextensive formalism [14]. This new approach will allow us to
understand open systems from a nonextensive point of view. We will
illustrate this by solving the model of the Van der Pol Oscilator in
the presence of friction and energy pumping. One important aspect of
this work is that this generalization can only be achieved using the
ordinary probability distribution rather than the escort
distribution.

\noindent \qquad In Section II, we present RE within the context of
ordinary statistics and show its relation to KL entropy. In Section
III, we study RE within the framework of nonextensive formalism. The
relation of NRE and the associated relative entropies is discussed
in Section IV. The ordinary and generalized S theorems is applied to
the Van der Pol oscillator in Section V. The results are summarized
in Section VI.
\bigskip
\bigskip
\bigskip
\bigskip
\bigskip
\bigskip
\bigskip
\bigskip
\bigskip
\bigskip
\bigskip
\bigskip
\bigskip
\bigskip
\bigskip
\bigskip
\bigskip
\bigskip
\bigskip
\bigskip
\bigskip
\bigskip

\section{Renormalized Entropy and Kullback-Leibler Entropy}

We begin by supposing two different probability distributions i.e.
p=\{p$_{i}\}$ and r=\{r$_{i}\}$. These distributions refer to the
state of a physical system with different control parameters, for
example [1]. Both of them are normalized to unity i.e.,

\begin{equation}
\sum_{i}p_{i}=\sum_{i}r_{i}=1.
\end{equation}

Comparing these two states in order to understand which one is more
ordered than the other by using the associated BG entropies is in
general not possible since the energies in both states may be
different, as in the case of Van der Pol oscillator [2]. However, it
will still be possible to calculate them. For example, for a
probability distribution p, we can calculate the corresponding BG
entropy

\begin{equation}
S(p)=-\sum_{i}p_{i}\ln p_{i},
\end{equation}

where units are chosen such that the Boltzmann constant $k$ is taken
to be equal to one. Following Ref. [1], we introduce the effective
Hamiltonian H$_{eff}$ of the system as

\begin{equation}
H_{eff}=-\ln r.
\end{equation}

The escort probability corresponding to the distribution
r=\{r$_{i}\}$ is given by

\begin{equation}
\widetilde{r}_{i}=\frac{r_{i}^{\beta }}{C}.
\end{equation}

where $\beta$ is a positive integer. Next, we renormalize energies
by setting

\begin{equation}
\left\langle H_{eff}\right\rangle ^{(0)}=\left\langle
H_{eff}\right\rangle ^{(1)},
\end{equation}

where superscripts denote the different states. Using Eq. (3), it
can be written as

\begin{equation}
\sum_{i}\widetilde{r}_{i}\ln r_{i}=\sum_{i}p_{i}\ln r_{i}.
\end{equation}

The ordinary renormalized entropy is defined as

\begin{equation}
R(p\Vert\widetilde{r})\equiv S(p)-S(\widetilde{r}).
\end{equation}

By explicitly substituting the Shannon entropies given by Eq. (2),
we obtain

\begin{eqnarray}
R(p\Vert\widetilde{r}) &\equiv &S(p)-S(\widetilde{r}) \\
&=&-\sum_{i}p_{i}\ln p_{i}+\sum_{i}\widetilde{r}_{i}\ln
\widetilde{r}_{i}.
\end{eqnarray}

Using Eq. (4) for the second term on the right hand side of the
equation, we get

\begin{equation}
R(p\Vert\widetilde{r})=-\sum_{i}p_{i}\ln p_{i}+\beta \sum_{i}\widetilde{r}%
_{i}\ln r_{i}-\sum_{i}\widetilde{r}_{i}\ln C.
\end{equation}

Using Eq. (6) for the second term on the right hand side of the
equation above,

\begin{equation}
R(p\Vert\widetilde{r})=-\sum_{i}p_{i}\ln p_{i}+\beta
\sum_{i}p_{i}\ln r_{i}-\sum_{i}\widetilde{r}_{i}\ln C.
\end{equation}

After a little algebra and using the normalization condition, we see
that

\begin{equation}
R(p\Vert\widetilde{r})=-\sum_{i}p_{i}\ln (p_{i}/\widetilde{r}_{i}).
\end{equation}

Comparing Eq. (12) with KL entropy [10] which is given by the
following equation

\begin{equation}
K[p\Vert r]\equiv \sum_{i}p_{i}\ln (p_{i}/r_{i}),
\end{equation}

we see the relation observed by Quiroga et al. [9], i.e.

\begin{equation}
R(p\Vert\widetilde{r})=-\sum_{i}p_{i}\ln
(p_{i}/\widetilde{r}_{i})=-K[p\Vert \widetilde{r}].
\end{equation}

This final result shows us that RE and K-L entropy is related to one
another by a factor of minus one. In other words, one can use RE or
K-L entropies in order to study self-organization once one employs
the escort distribution and renormalization of mean energy values.

\section{Renormalized Entropy and Tsallis Entropy}

\qquad A nonextensive generalization of the standard Boltzmann-Gibbs
(BG) entropy has been proposed by C. Tsallis in 1988 [15-18]. The
nonextensive formalism has been used successfully to investigate
earthquakes [19], models of fracture roughness [20], entropy
production [21], Ising chains [22] and climatological models [23].
This new definition of entropy is given by

\begin{equation}
S_{q}(p)=\frac{\sum_{i}p_{i}^{q}-1}{1-q},
\end{equation}

where p$_{\text{i }}$ is the probability of the system in the ith
microstate, W is \ the total number of the configurations of the
system. The entropic index q is a real number, which characterizes
the degree of nonextensivity as can be seen from the following
pseudo-additivity rule:

\begin{equation}
S_{q}(A+B)/k=[S_{q}(A)/k]+[S_{q}(B)/k]+(1-q)[S_{q}(A)/k][S_{q}(B)/k],
\end{equation}
where A and B are two independent systems i.e., p$_{ij}$(A+B)$=$p$_{i}(A)$p$%
_{j}(B)$. As q$\rightarrow 1$, the nonextensive entropy definition
in Eq. (15) becomes

\begin{equation}
S_{q\rightarrow 1}=-\sum_{i=1}^{W}p_{i}\ln p_{i},
\end{equation}
which is the usual BG entropy already given by Eq. (2). This means
that the definition of nonextensive entropy contains BG statistics
as a special case. The cases q
\mbox{$<$}%
1, q
\mbox{$>$}%
1 and $q=1$ correspond to superextensivity, subextensivity and
extensivity, respectively. We define the effective Hamiltonian for
this case as

\begin{equation}
H_{eff}=\frac{r^{q-1}-1}{1-q}=\ln _{q}(1/r),
\end{equation}

where q-logarithm is defined as

\begin{equation}
\ln _{q}(x)=\frac{x^{1-q}-1}{1-q}.
\end{equation}

Note that as $q$ approaches to 1, the effective Hamiltonian given by
Eq. (18) becomes identical to the one given by Eq. (3) in BG case
since q-logarithm given by Eq. (19) becomes equal to natural
logarithm in this limiting case. Setting the mean energy of the two
states to be equal to one another as we have done before is
tantamount to writing

\begin{equation}
\sum_{k}\frac{r_{k}^{q-1}-1}{1-q}\widetilde{r}_{k}=\sum_{k}\frac{%
r_{k}^{q-1}-1}{1-q}p_{k}.
\end{equation}

Using the fact that the probability distributions are normalized, we
can write Eq. (20) as

\begin{equation}
\sum_{k}r_{k}^{q-1}\widetilde{r}_{k}=\sum_{k}r_{k}^{q-1}p_{k}.
\end{equation}

Using the definition of RE given by Eq. (7), we can form the NRE

\begin{equation}
R_{q}(p\Vert\widetilde{r})=-[\frac{1}{(q-1)}(\sum_{k}p_{k}^{q}-\sum_{k}\widetilde{r}%
_{k}^{q})].
\end{equation}

We can rewrite the equation above as

\bigskip
\begin{equation}
R_{q}(p\Vert\widetilde{r})=-[\frac{1}{(q-1)}(\sum_{k}p_{k}^{q}-\sum_{k}\widetilde{r}%
_{k}^{q}+(q-1)\sum_{k}\widetilde{r}_{k}^{q}-(q-1)\sum_{k}\widetilde{r}%
_{k}^{q})].
\end{equation}

Using the ordinary probability definition by taking $\beta=1$ in Eq.
(4) i.e.,
\begin{equation}
\widetilde{r}_{k}\equiv \frac{r_{k}}{C}.
\end{equation}

we can write

\begin{equation}
\sum_{k}\widetilde{r}_{k}^{q}=\sum_{k}\frac{r_{k}}{C}\frac{r_{k}^{q-1}}{%
C^{q-1}}=\frac{1}{C^{q-1}}\sum \widetilde{r}_{k}r_{k}^{q-1}.
\end{equation}

Using Eq. (21), we see that

\begin{equation}
\sum_{k}\widetilde{r}_{k}^{q}=\frac{1}{C^{q-1}}\sum_{k}p_{k}r_{k}^{q-1}.
\end{equation}

which is equal to

\begin{equation}
\sum_{k}\widetilde{r}_{k}^{q}=\sum_{k}p_{k}\widetilde{r}_{k}^{q-1}.
\end{equation}

Using the relation above in Eq. (23) for the last two terms, we
obtain

\begin{equation}
R_{q}(p\Vert\widetilde{r})=-(\frac{\sum_{k}p_{k}^{q}}{q-1}+\sum_{k}\widetilde{r}%
_{k}^{q}-\frac{1}{q-1}\sum_{k}p_{k}\widetilde{r}_{k}^{q-1}-\sum_{k}p_{k}%
\widetilde{r}_{k}^{q-1}).
\end{equation}

Before proceeding with an analysis of this explicit form of NRE, we
need to see the two different expressions of relative entropy in
nonextensive formalism.

\section{Nonextensive Renormalized Entropy and Nonextensive Relative Entropies}

In nonextensive formalism, we have two different expressions of
relative entropy. The first one is of Bregman type [24-26] and is
given by

\begin{equation}
K_{q}[p\Vert r]=\frac{\sum_{k}p_{k}^{q}}{q-1}+\sum_{k}{r}%
_{k}^{q}-\frac{1}{q-1}\sum_{k}p_{k}{r}_{k}^{q-1}-\sum_{k}p_{k}%
{r}_{k}^{q-1},
\end{equation}

whereas the second one is of Csisz\'{a}r type [27-31] and reads

\begin{equation}
I_{q}[p\Vert r]=\frac{1}{1-q}[1-\sum%
\limits_{k}p_{k}^{q}r_{k}^{1-q}].
\end{equation}

These two forms of relative entropies have also been used in quantum
theoretical framework in order to study second law of thermodynamics
and purity of states in quantum information related contexts
[32-35]. In order to see the close connection between these relative
entropies and K-L entropy, note that it can be written in the
following form

\begin{equation}
K[p\Vert r]=\frac{d}{dx}\sum\limits_{i}(p_{i})^{x}(r_{i})^{1-x}\mid
_{x\rightarrow 1}.
\end{equation}

This form is preserved exactly if one uses Jackson $q$-differential
operator [32, 36-38] instead of ordinary differential operator above
i.e.,

\bigskip
\begin{equation}
I_{q}[p\Vert r]=D_{q}\sum\limits_{i}(p_{i})^{x}(r_{i})^{1-x}\mid
_{x\rightarrow 1},
\end{equation}

where Jackson $q$-differential operator [36] is defined as

\begin{equation}
D_{q}f(x)=[f(qx)-f(x)]/[x(q-1)].
\end{equation}

However, there is no such simple correspondence in the case of
relative entropy of Bregman type [39]. Yet, one can still write

\bigskip
\begin{equation}
K[p\Vert r]=\frac{dG(x)}{dx}\mid _{x\rightarrow 1},
\end{equation}

where the function $G(x)$ is given by

\begin{equation}
G(x)=\frac{\sum_{k}p_{k}^{x}}{x}-\frac{\sum_{k}r_{k}^{x}}{x}%
-\sum_{k}p_{k}r_{k}^{x-1}+\sum_{k}r_{k}^{x}.
\end{equation}

Then, we have the following relation between the relative entropy of
Bregman type and the function $G(x)$

\begin{equation}
K_{q}[p\Vert r]=qD_{q}G(x)\mid _{x\rightarrow 1}.
\end{equation}

Finally, we note that both forms of nonextensive relative entropies
are non-negative and equal to zero if and only if two distributions
are equal to one another preserving these properties shared by K-L
entropy [26, 39]. Both of these relative entropies become K-L
entropy as the parameter $q$ approaches 1.

In order to see the relation between RE and nonextensive formalism,
we compare Eqs. (28), (29) and (30), to see that

\begin{equation}
R_{q}(p\Vert\widetilde{r})=-K_{q}[p\Vert \widetilde{r}].
\end{equation}

The nonextensive renormalized entropy cannot be written in terms of
relative entropy of Csisz\'{a}r kind given by Eq. (30). In other
words, it is possible to obtain an equation in nonextensive case
similar to Eq. (14) in ordinary case, showing that nonextensive
renormalized entropy is nothing but the nonextensive relative
entropy multiplied by a factor of minus one only by adopting the
relative entropy of Bregman type given by Eq. (29). There are three
important points to be noted. First, we have used Eq. (27) in order
to reach Eq. (28). On the other hand, this equation itself is based
on Eq. (21) which is the renormalization relation in nonextensive
formalism corresponding to Eq. (6) in ordinary case. This simply
shows the necessity of energy renormalization even in nonextensive
formalism. Second point is the use of Eq. (24). This is nothing but
a definition of an ordinary probability distribution with a
normalization constant $C$. This is very different than BG case
where we have used Eq. (4) i.e., escort distribution. In other
words, we are forced to use ordinary probability in nonextensive
formalism in the same way we are forced to use escort distribution
within BG statistics (and with its corresponding relative entropy
which is K-L entropy). Lastly, we have two different relative
entropy definitions but we need to use only one of them in order to
define the nonextensive version of renormalized entropy. In order to
assess the importance of the relation given by Eq. (37), we need to
understand one subtle point: In Ref. [39], it has been shown that
relative entropy of Bregman type is associated with the ordinary
constraint, whereas relative entropy of Csisz\'{a}r type is the one
associated with the escort distribution. This shows that the use of
ordinary probability and the nonextensive relative entropy
associated with ordinary constraint is enough to study
self-organization in nonextensive formalism if one wants to follow
Klimontovich's recipe.

\section{APPLICATIONS}

In this Section, we present first the application of the ordinary
renormalized entropy and then nonextensive renormalized entropy to
the Van der Pol oscillator and show that Klimontovich's S theorem is
satisfied in both cases resulting negative renormalized entropy
values for all of the involved parameters.

\subsection{Renormalized Entropy and the Van der Pol Oscillator}

Now, we apply the ideas explained in the previous Sections to the
case of the Van der Pol oscillator [2]. The equation for this case
in the presence of a Langevin source can be given as

\begin{equation}
\frac{dx}{dt}=v,\frac{dv}{dt}+(a+bE)v+\omega _{0}^{2}x=y(t),
\end{equation}

where $\omega _{0}$ is the eigenfrequeny, b is the nonlinear
friction coefficient. The term $a$ can be written in terms of two
other parameters i.e.,

\begin{equation}
a=\gamma -a_{f},
\end{equation}

where $\gamma$ is the coefficient of linear friction and $a_{f}$ is
the feedback coefficient. The term E is nothing but the energy of
oscillation where mass term is taken to be equal to unity.

\begin{equation}
E=\frac{1}{2}(v^{2}+\omega _{0}^{2}x^{2}).
\end{equation}

The random Langevin source term can be defined by the following
equations.
\begin{equation}
\left\langle y(t)\right\rangle =0,\left\langle y(t)y(t)^{\prime
}\right\rangle =2D\delta (t-t^{\prime }),
\end{equation}

where the the intensity of random source D is a given positive
constant and not connected with the temperature via the Einstein
formula. For the case when the following conditions hold

\begin{equation}
\gamma ,\left\vert a\right\vert ,b\left\langle E\right\rangle \ll
\omega _{0},
\end{equation}

one can write the following Fokker-Planck equation for the
distribution function $f(E,t)$:

\begin{equation}
\frac{\partial f(E,t)}{\partial t}=\frac{\partial }{\partial E}(DE\frac{%
\partial f}{\partial E})+\frac{\partial }{\partial E}[(a+bE)Ef].
\end{equation}

The solution to the equation above for the stationary case is given
by

\begin{equation}
f_{0}(E)=C\exp (-\frac{aE+\frac{1}{2}bE^{2}}{D}).
\end{equation}

where $C$ is the normalization constant. The state of equilibrium
corresponds to the case when the feedback parameter $a_{f}$ is equal
to zero. Then, the corresponding distribution function, adopting the
same notation in previous Section, becomes

\begin{equation}
r(E)=C\exp (-\frac{\gamma E+\frac{1}{2}bE^{2}}{D}).
\end{equation}

Assuming the good oscillator condition [14] i.e.,

\begin{equation}
b\left\langle E\right\rangle /\gamma \sim Db/\gamma ^{2}\ll 1,
\end{equation}

we obtain the following equilibrium distribution function

\begin{equation}
r(E)=\frac{\gamma }{D}\exp (-\frac{\gamma E}{D}).
\end{equation}

Using the formula $\left\langle E\right\rangle =\int dEf(E,t)E.$ in
order to calculate the average energy and Eq. (2) in the continuous
case, we obtain for the entropy

\begin{equation}
S(r)=\ln (\frac{D}{\gamma })+1.
\end{equation}

and the energy

\begin{equation}
\left\langle E\right\rangle ^{(1)}=\frac{D}{\gamma }.
\end{equation}

respectively. The threshold of generation is defined as the state
when feedback parameter $a_{f}$ is equal to $\gamma$. Then,
according to Eq. (39), $a$=0. Therefore, the distribution function
for this case can be written as

\begin{equation}
p(E)=\sqrt{\frac{2b}{\pi D}}\exp (-\frac{bE^{2}}{2D}).
\end{equation}

The corresponding entropy and energy values are calculated as

\begin{equation}
S(p)=\ln (\sqrt{\frac{\pi D}{2b}})+\frac{1}{2}.
\end{equation}

and

\begin{equation}
\left\langle E\right\rangle ^{(2)}=\sqrt{\frac{2D}{\pi b}}.
\end{equation}

respectively. So, we have two entropy values corresponding to two
distinct values of the control parameter $a_{f}$. Although we would
expect the equilibrium entropy to be greater than the
off-equilibrium case which is characterized by nonzero control
parameter, wee see that this is not the case since the entropy given
by Eq. (48) is less than the value given by Eq. (51) if we take into
account also Eq. (46). In order to solve this problem, we
renormalize these two states so that their energies will be taken to
be equal to one another so that we will have new intensity for the
random force of the equilibrium state. Therefore, writing

\begin{equation}
\left\langle \widetilde{E}\right\rangle ^{(1)}=\left\langle \widetilde{E}%
\right\rangle ^{(2)}=\sqrt{\frac{2D}{\pi b}},
\end{equation}

we obtain the new intensity of the random force as

\begin{equation}
\widetilde{D}^{(1)}=\gamma \sqrt{\frac{2D}{\pi b}}.
\end{equation}

Substitution of this new expression into Eq. (48) gives us the new
renormalized entropy

\begin{equation}
S(\widetilde{r})=\ln (\sqrt{\frac{2D}{\pi b}})+1.
\end{equation}

It is easily seen that the renormalized equilibrium entropy in Eq.
(55) is now greater than the off-equilibrium entropy given by Eq.
(51).

Note that another way to see this is directly to use Klimontovich's
S theorem which states that renormalized entropy defined by Eq. (7)
decreases in this case. This is tantamount to say that the
difference of the entropy given by Eq. (51) and Eq. (55) is less
than zero. If we calculate it explicitly using Eqs. (51) and (55),
we see that

\begin{equation}
R_{q}(p\Vert\widetilde{r})\equiv S(p)-S(\widetilde{r})=-0.05<0.
\end{equation}

This simple observation will be important when we look at the same
problem from the nonextensive point of view.

\subsection{Nonextensive Renormalized Entropy and the Van der Pol Oscillator}

We now apply the abstract formalism, which has been developed in
Sections IV and V to the problem of the Van der Pol oscillator. We
have already studied this example in the context of ordinary
renormalized entropy. The only difference in treatment will then be
the adoption of Tsallis entropy instead of ordinary Boltzmann-Gibbs
entropy. We will assume the underlying mechanics does not change so
that we will use the same distribution functions. Therefore, we
begin by combining Eqs. (29) and (37) in order to write

\begin{equation}
R_{q}(p\Vert\widetilde{r})=S_{q}(p)-S_{q}(\widetilde{r})=\frac{1}{1-q}\int\limits_{0}^{%
\infty }dEp^{q}-\int\limits_{0}^{\infty
}dE\widetilde{r}^{q}+\frac{1}{q-1}\int\limits_{0}^{\infty
}dEp\widetilde{r}^{q-1}+\int\limits_{0}^{\infty
}dEp\widetilde{r}^{q-1}.
\end{equation}

The nonextensive generalization of the renormalized entropy requires
the use of ordinary probability. This means that the average energy
values we have already obtained in the context of ordinary
renormalized entropy are still valid in the nonextensive context.
Therefore, we will still use Eqs. (49) and (52) even though we will
adopt Tsallis entropy. Using the previously obtained distribution
functions i.e., Eqs. (47) and (50) in Eq. (57) above, we can
calculate the first integral on the right hand side as follows

\begin{equation}
\int_{0}^{\infty }dEp^{q}=\int_{0}^{\infty }dE(\frac{2b}{\pi D}%
)^{q/2}e^{-bqE^{2}/2D}=(\frac{2b}{\pi D})^{q/2}(\frac{D\pi
}{2bq})^{1/2}.
\end{equation}

The integral appearing in the second term can be calculated in a
similar manner but by using the renormalized equilibrium
distribution

\begin{equation}
\int_{0}^{\infty }dE\widetilde{r}^{q}=\int_{0}^{\infty }dE(\frac{\pi b}{2D}%
)^{q/2}e^{-\sqrt{\frac{\pi b}{2D}}qE}=\frac{1}{q}(\frac{\pi b}{2D}%
)^{(q-1)/2}.
\end{equation}

The only integral to be solved in Eq. (57) is then the following

\begin{equation}
\int_{0}^{\infty }dEp\widetilde{r}^{q-1}=\int_{0}^{\infty }dE(\frac{2b}{\pi D%
})^{1/2}(\frac{\pi b}{2D})^{(q-1)/2}e^{-(q-1)(\frac{\pi b}{2D})^{1/2}E}e^{-%
\frac{bE^{2}}{2D}}.
\end{equation}

This integral can be rearranged as

\begin{equation}
\int_{0}^{\infty }dE(\frac{2b}{\pi D})^{1/2}(\frac{\pi b}{2D}%
)^{(q-1)/2}e^{-(q-1)(\frac{\pi b}{2D})^{1/2}E}e^{-\frac{bE^{2}}{2D}}=\frac{2%
}{\pi }(\frac{\pi b}{2D})^{q/2}\int_{0}^{\infty }dE\exp [-(q-1)(\frac{\pi b}{%
2D})^{1/2}E-\frac{bE^{2}}{2D}].
\end{equation}

The integral above can be solved by the method of completing the
squares. Writing

\begin{equation}
\exp [-(q-1)(\frac{\pi b}{2D})^{1/2}E-\frac{bE^{2}}{2D}]=-\frac{b}{2D}%
[E+(q-1)(\frac{\pi D}{2b})^{1/2}]^{2}+(q-1)^{2}\frac{\pi }{4}
\end{equation}

and doing the following substitution

\begin{equation}
\sqrt{\frac{b}{2D}}[E+(q-1)(\frac{\pi D}{2b})^{1/2}]=z,
\end{equation}

we get

\begin{equation}
\int_{0}^{\infty }dE\exp [-(q-1)(\frac{\pi b}{2D})^{1/2}E-\frac{bE^{2}}{2D}%
]=e^{(q-1)^{2}\frac{\pi }{4}}\sqrt{\frac{\pi D}{2b}}{erf}c(\alpha ).
\end{equation}

where $\alpha$ is given by

\begin{equation}
\alpha =\frac{\sqrt{\pi }}{2}(q-1)
\end{equation}

The complementary error function $erfc(x)$ is defined as

\begin{equation}
erfc(x)=1-erf(x)=\frac{2}{\sqrt{\pi }}\int\limits_{x}^{\infty
}e^{-t^{2}}dt,
\end{equation}

where the error function $erf(x)$ is

\begin{equation}
erf(x)=\frac{2}{\sqrt{\pi }}\int\limits_{0}^{x}e^{-t^{2}}dt.
\end{equation}

Combining Eqs. (61), (64) and (65), we can write

\begin{equation}
\int_{0}^{\infty }dE(\frac{2b}{\pi D})^{1/2}(\frac{\pi b}{2D}%
)^{(q-1)/2}e^{-(q-1)(\frac{\pi b}{2D})^{1/2}E}e^{-\frac{bE^{2}}{2D}}=\frac{2%
}{\pi }(\frac{\pi b}{2D})^{q/2}e^{(q-1)^{2}\frac{\pi }{4}}\sqrt{\frac{\pi D}{%
2b}}{erf}c(\frac{\sqrt{\pi }}{2}(q-1)).
\end{equation}

Using Eqs. (58), (59) and (68), we finally obtain the nonextensive
renormalized entropy as follows

\begin{eqnarray}
R_{q}(p\Vert\widetilde{r})=S_{q}(p)-S_{q}(\widetilde{r})=\frac{1}{(1-q)}((\frac{2b}{\pi D}%
)^{q/2}\sqrt{\frac{\pi D}{2bq}})-\frac{1}{q}(\frac{\pi b}{2D})^{(q-1)/2}+\nonumber \\%
(\frac{q}{q-1})\frac{2%
}{\pi }(\frac{\pi b}{2D})^{q/2}e^{(q-1)^{2}\frac{\pi }{4}}\sqrt{\frac{\pi D}{%
2b}}{erf}c(\frac{\sqrt{\pi }}{2}(q-1)).
\end{eqnarray}

The relation given by Eq. (37) guarantees that the NRE like its
ordinary counterpart will take only negative values since the
relative entropy of Bregman type on the right hand side of Eq. (37)
is positive definite (it only becomes zero when the two probability
distributions are equal to one another) [26]. Due to the minus sign
in front of it, the NRE is seen to be negative definite for all
values of the parameters $D$, $b$ and $q$. In Figs. 1 and 2, we plot
NRE for some particular values of the intensity of the random source
$D$ and nonlinear friction coefficient $b$ as a function of the
nonextensivity parameter $q$. In all these cases, NRE takes only
negative values thereby ordering the entropies. The NRE
$R_{q}(p\Vert\widetilde{r})$ attains the value $-0.05$ as the
nonextensivity index $q$ becomes 1 independent of the values of $D$
and $b$. Note that this is exactly the value one obtains by using
ordinary RE.

\section{CONCLUSIONS}

It is known that BG entropy is inadequate to handle the systems
which depend on a control parameter [1-4]. In these types of
systems, one has different entropy and energy values corresponding
to different values of the control parameter and the maximum entropy
value is not necessarily attained by the equilibrium distribution.
The solution for this problem has been provided by Klimontovich's S
theorem which is based on the joint use of energy renormalization
and escort distribution. This defect is shared by Tsallis entropy
since BG entropy is a particular case of Tsallis entropy. In this
work, we have generalized S theorem in the context of nonextensive
formalism and showed that this is possible only when one adopts
ordinary probability distribution. If one uses ordinary distribution
and the associated relative entropy together with the energy
renormalization condition, one obtains a nonextensive renormalized
entropy that is a generalization of Klimontovich's S theorem.
Therefore, what can be achieved in the ordinary BG entropy with
energy renormalization and escort distribution in the context of S
theorem can be achieved through the use of Tsallis entropy,
renormalization of energy and ordinary probability distribution. As
a result, we conclude that the use of ordinary probability in
nonextensive formalism must not be underestimated since it can be
the only form of probability distribution needed in some cases such
as the generalization of Klimontovich's S theorem. We have also
applied the nonextensive generalization of S theorem to the Van der
Pol oscillator and have shown that the value of nonextensive
renormalized entropy is negative definite independent of all the
parameters involved and attains $-0.05$ as the nonextensivity index
$q$ becomes 1 exactly giving the numerical value which one would
obtain by using ordinary renormalized entropy. This new nonextensive
measure of complexity could be used in the analysis of logistic maps
and heart rates as it has been the case with the ordinary
renormalized entropy [1, 6, 7].

\section{ACKNOWLEDGEMENTS}
We thank Donald H. Kobe for fruitful discussions.

\bigskip
\bigskip
\bigskip
\bigskip
\bigskip
\bigskip
\bigskip
\bigskip
\bigskip
\bigskip
\bigskip
\bigskip
\bigskip
\bigskip
\bigskip
\bigskip
\bigskip
\bigskip
\bigskip
\bigskip

\begin{figure}
\includegraphics[width=15cm, height=10cm]{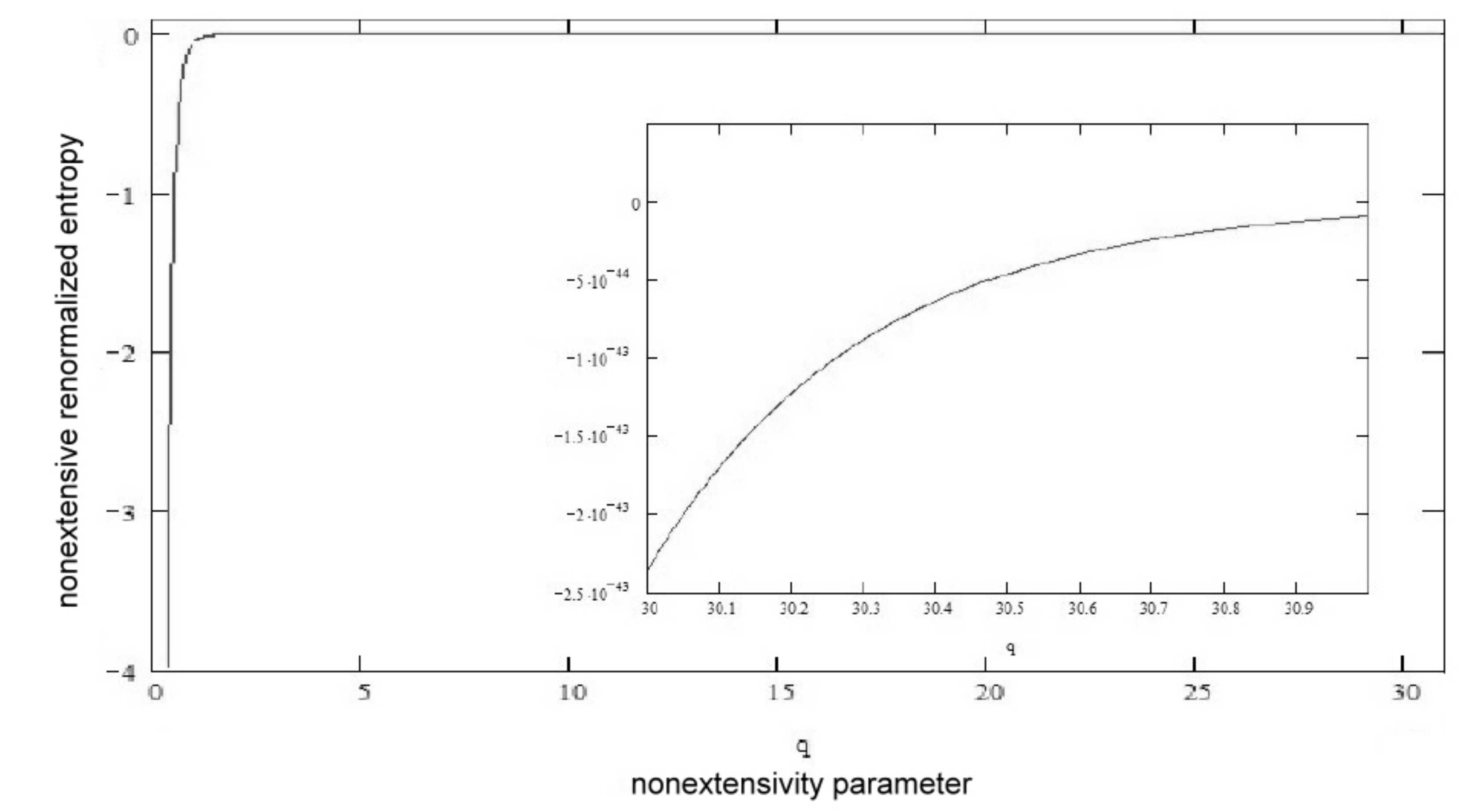}
\caption{\label{fig1} The renormalized nonextensive entropy versus
nonextensivity parameter $q$ where the intensity of the random
source $D$=50 and nonlinear friction coefficient $b$=0.05}
\end{figure}

\begin{figure}
\includegraphics[width=15cm, height=10cm]{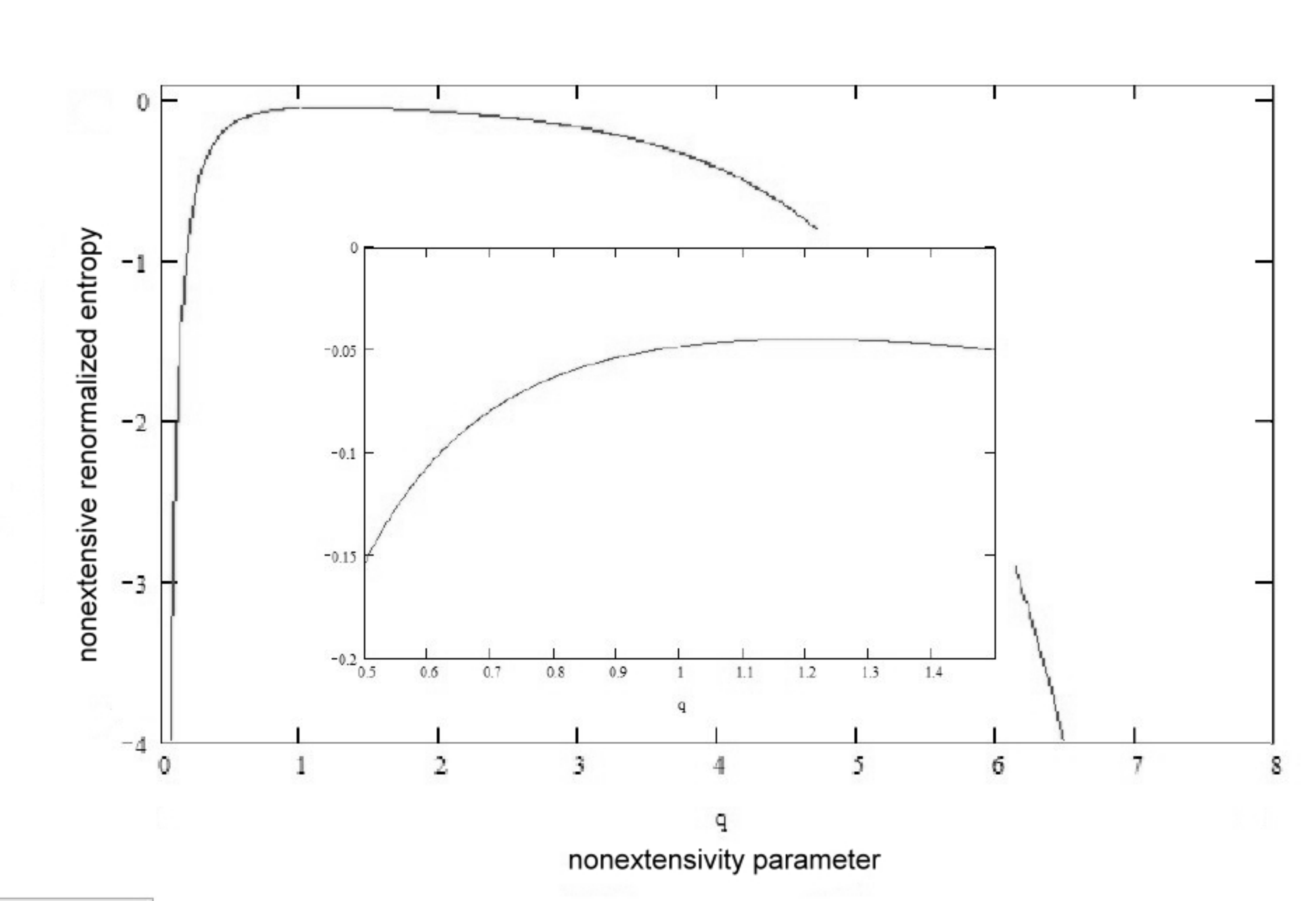}
\caption{\label{fig2} The renormalized nonextensive entropy versus
nonextensivity parameter $q$ where the intensity of the random
source $D$=5 and nonlinear friction coefficient $b$=20}
\end{figure}


\begin{references}

\bibitem{R1}  P. Saparin, A. Witt, J. Kurths, V. Anischenko, Chaos, Solitons and Fractals{\bf 4}, 1907 (1994).

\bibitem{R2}  Yu. L. Klimontovich, Physica A {\bf 142}, 390 (1987).

\bibitem{R3}  Yu. L. Klimontovich, Chaos, Solitons and Fractals {\bf 5}, 1985 (1994).

\bibitem{R4}  Yu. L. Klimontovich, Turbulent Motion and the Structure of Chaos: A New Approach to the Statistical Theory of Open System, Kluwer Academic Publishers, Dordrecht, 1991.

\bibitem{R5}  Yu. L. Klimontovich, Z. Phys. B {\bf 66}, 125 (1987).

\bibitem{R6}  J. Kurths \textit{et al.}, Chaos {\bf 5}, 88 (1995).

\bibitem{R7}  A. Voss \textit{et al.}, Cardiovasc. Res. {\bf 31}, 419 (1996).

\bibitem{R8}  K. Kopitzki, P. C. Warnke, J. Timmer, Phys. Rev. E {\bf 58}, 4859 (1998).

\bibitem{R9}  R. Q. Quiroga, J. Arnold, K. Lehnertz, P. Grassberger, Phys. Rev. E {\bf 62}, 8380 (2000).

\bibitem{R10} R. Gray, Entropy and Information Entropy, Springer-Verlag, New York, 1990.

\bibitem{R11} C. Beck, F. Schl\"{o}gl, Thermodynamics of Chaotic Systems, Cambridge University Press, Cambridge, 1993.

\bibitem{R12}  K. Kopitzki \textit{et al.}, Phys. Rev. E {\bf 66}, 043902 (2002).

\bibitem{R13}  R. Q. Quiroga, J. Arnold, K. Lehnertz, P. Grassberger, Phys. Rev. E {\bf 66}, 043903 (2002).

\bibitem{R14}  C. Tsallis, J.Stat. Phys. {\bf 52}, 479 (1988). \

\bibitem{R15}  Yu. L. Klimontovich, Statistical Physics, Harwood Academic Publishers, New York, 1986.

\bibitem{R16}  C. Tsallis, in: New Trends in Magnetism, Magnetic Materials
and their Applications, eds. J.L.Mor\'{a}n-Lopez and J.M.
S\'{a}nchez (Plenum Press, New York, 1994), p.451.\

\bibitem{R17}  C. Tsallis, Some comments on Boltzmann-Gibbs statistical
mechanics, Chaos, Solitons and Fractals{\bf \ 6}, 539 (1995).

\bibitem{R18}  E.M.F. Curado and C. Tsallis, J. Phys. A {\bf 24} (1991) L69;
Corrigenda: J. Phys. A {\bf 24} (1991) 3187; {\bf 25}, 1019 (1992).

\bibitem{R19}  R. Silva, G. S. Fran\c{c}a, C. S. Vilar, J. S. Alcaniz, Phys. Rev. E {\bf 73}, 026102 (2006).

\bibitem{R20}  S. Nadarajah, Samuel Kotz, Physics Letters A {\bf 359}, 577 (2006).

\bibitem{R21}  Sumiyoshi Abe, Yutaka Nakada, Phys. Rev. E {\bf 74}, 021120 (2006).

\bibitem{R22}  R. F. S. Andrade, S. T. R. Pinho, Phys. Rev. E {\bf 71}, 026126 (2005).

\bibitem{R23}  M. Ausloos, F. Betroni, Physica A {\bf 373}, 721 (2007).

\bibitem{R24}  L. M. Bregman, USSR Comput. Math. Math. Phys. {\bf 7}, 200 (1967).

\bibitem{R25}  T. D. Frank, Nonlinear Fokker-Planck Equations: Fundamentals and Applications, Springer Verlag, Berlin, 2005.

\bibitem{R26}  Jan Naudts, Rev. Math. Phys. {\bf 16}, 809 (2004).

\bibitem{R27}  I. Csisz\'{a}r, Period. Math. Hung. {\bf 2}, 191 (1972).

\bibitem{R28}  C.Tsallis, Phys.Rev.E {\bf 58}, 1442 (1998).

\bibitem{R29}  M.Shiino, J.Phys.Soc.Japan {\bf 67} 3658 (1998).

\bibitem{R30}  L.Borland, A.R.Plastino and C.Tsallis, J.Math.Phys. {\bf 39}, 6490 (1998); Erratum: {\bf 40} 2196 (1999).

\bibitem{R31}  S. Furuichi, K.Yanagi, K.Kuriyama, J. of Math. Phys. {\bf 45}, 4868-4877 (2004).

\bibitem{R32}  S.Abe, Phys.Rev.A {\bf 68} 032302 (2003).

\bibitem{R33}  S.Abe and A.K.Rajagopal, Phys.Rev.Lett. {\bf 91}, 120601(2003).

\bibitem{R34}  G. B. Bagci, A. Arda, R. Sever, Int. J. Mod. Phys. {\bf 20}, 2085 (2006).

\bibitem{R35}  G. B. Bagci, A. Arda, R. Sever, Mod. Phys. Lett. B, in press.

\bibitem{R36}  F. Jackson, Mess. Math. {\bf 38}, 57 (1909); Quart. J. Pure Appl.
Math. {\bf 41}, 193 (1910).

\bibitem{R37}  S. Abe, Phys. Lett. A {\bf 224}, 326 (1997).

\bibitem{R38}  S. Abe, Phys. Lett. A {\bf 244}, 229 (1998).

\bibitem{R39}  S. Abe, G. B. Bagci, Phys. Rev. E {\bf 71}, 016139 (2005).


\end{references}
\end{document}